\documentclass{elsart}
\usepackage{graphicx}

\newcommand{\lsim}{\mbox{\raisebox{-1.ex}{$\stackrel{<}{\sim}$}}}
\newcommand{\gsim}{\mbox{\raisebox{-1.ex}{$\stackrel{>}{\sim}$}}}
\newcommand{\vct}[1]{\mbox{\boldmath${#1}$}}

\begin{document}
\begin{frontmatter}

\title{Production cross sections of $\gamma$-rays, electrons, and positrons
 in p-p collisions
}

T.\ Shibata$^1$\footnote{Corresponding author.
Tel/fax: +81 042 331 6515. \\
 {\it E-mail address:} shibata@phys.aoyama.ac.jp}, 
Y.\ Ohira$^1$, K.\ Kohri$^2$, and R.\ Yamazaki$^1$

{\footnotesize $^{1}$
Department of Physics and Mathematics, Aoyama-Gakuin University,
Kanagawa 252-5258, Japan}

{\footnotesize $^{2}$
KEK Theory Center and the Graduate University for Advanced Studies
 (Sokendai) 1-1 Oho,  
 Tsukuba 305-0801, Japan}



\begin{abstract}
Because the production cross sections of $\gamma$ rays, electrons, and positrons (e$^\pm$) made in
 p-p collisions, $\sigma_{pp\rightarrow \gamma}$ and $\sigma_{pp\rightarrow {e}^\pm}$, respectively,
are kinematically equivalent with respect to the parent pion-production
cross section $\sigma_{pp\rightarrow \pi}$,  we obtain   
$\sigma_{pp\rightarrow {e}^\pm}$ directly from the machine data on 
$\sigma_{pp\rightarrow \gamma}$.
 In Paper I (Sato et al.\ [1]), we give explicitly $\sigma_{pp\rightarrow 
\gamma}$, reproducing quite well the accelerator data with LHC, 
 namely $\sigma_{pp\rightarrow {e}^\pm}$ is applicable enough 
 over the wide energy range from GeV to 20\,PeV for projectile proton energy.
 We dicuss in detail the relation between the cross sections, 
 and present explicitly $\sigma_{pp\rightarrow {e}^\pm}$
that are valid into the PeV electron energy. 

\end{abstract}

\end{frontmatter}

\section{Introduction.}

Fluxes of cosmic-ray (CR) antimatter, 
namely  antiprotons ($\bar{\rm p}$'s) and positrons (e$^+$'s),
have been measured with balloon and satellite experiments.
Notable recent experiments in the higher-energy, $\gsim$ 100\,GeV, regime,  
include PAMELA [2,3] and AMS-02 [4].
 While the complete results
 of the latter experiment are not yet available,
the improved accuracy of new CR data motivates new studies of 
the underlying production cross sections and collision kinematics.

 Since the early reports around 1980 of the anomaly in the antiproton spectrum   [5-7], cosmic-ray physics has shown itself 
in a position to test
and potentially challenge
 conventional models in both particle physics and astrophysics. The  
antiproton anomaly in this experiment 
suffered, however, from limited sensitivity of the experiment, separation of
 atmospheric $\bar{\rm p}$'s, modulation effects, and low statistics.

The low-energy antiproton excess is not found
 in recent $\bar{\rm p}$ spectral data, for instance, 
BESS [8]
 and PAMELA [2], which gives data from $\sim 200$\,MeV  
 up to $\sim 100$\,GeV. 
 The  $\bar{\rm p}$ spectrum is adequately explained by
 the standard model for the production of $\bar{\rm p}$ in the Galaxy and 
 subsequent propagation to the
 Earth, within the uncertainties in the choice of model parameters.
 
 Earlier, the HEAT (High Energy Antimatter Telescope) experiment [9] 
reported  in 1995 a possible anomaly in the positron fraction. 
Connection of this excess to an explanation involving  
Weakly Interacting Massive Particles (WIMPs) has been suggested [10, 11].
DuVernois et al.\ [9] presented data indicating 
an  excess of e$^+$ in the higher energy region as compared to the
 expectation from the standard Galactic production and propagation models, 
and many antimatter searches have been performed since these early studies.

For instance, PAMELA [2, 12] recently reported with good statistics that
the positron fraction increases with increasing energy
above $\sim$\,4\,GeV 
 over the 0.1 -- 150\,GeV energy range. {\it Fermi}
 [13] also finds that the positron fraction in the  20-150\,GeV
is higher than expected from standard galactic propagations models.
Both results indicate that
 the positron fraction rises significantly above 4\,GeV up to 100\,GeV.

AMS-02 will, in the near future, confirm the positron excess with higher statistics and unprecedented
 precision over a wide energy range from
 0.5\,GeV -- 1\,TeV [14]. It is particularly
 interesting to see if they observe a peak of the positron fraction 
 around 350\,GeV; the first result seems to be reaching a 
plateau at these energies [4].

The aim of the  present paper is not to give an explanation 
for the anomaly in the positron spectrum, 
but rather to provide improved production cross sections for
 $\sigma_{pp\rightarrow s}
 (E_0, E_s)$ ($\lq\lq s$"\,$\equiv$\,$\gamma, \mbox{e}^\pm$, $\nu$) 
for decay product $s$ coming from pions produced in p-p collisions,
 and to present the kinematical relations between them. 
Here and in the following, 
$E_0$ and $E_s$ denote the laboratory frame (L.F.)
 kinetic energy of the projectile proton and the total energy of 
secondary {\it s}, respectively.

The parameterization for $\sigma_{pp\rightarrow \pi^0 \rightarrow \gamma}$ has
 been studied by many authors [15-21], giving convenient forms for practical 
applications to the galactic phenomena.
It is, however, experimentally very difficult 
to measure $\sigma_{pp\rightarrow \pi^\pm}$ for {\it energetic} charged 
 pions produced by p-p collision in the high energy region.
 In fact, no data on $\sigma_{pp\rightarrow \pi^\pm} 
(E_0, E_{\pi^\pm})$ are available above intersecting-storage-ring energies [22] with $E_0$\,$\sim$\,1\,TeV, before 
LHCf [23].
This is mainly due to the concentration of secondary products 
in the beam-fragmentation region that are too collimated to separate
in the detector, 
in addition to the difficulty in energy determination of 
 a single charged track with  energy larger than $\approx 1$\,TeV.

Previously, we have focused upon $\gamma$-ray production via $\pi^0$
instead of $\pi^\pm$, the energy determination
 of which is rather easy and reliable even in the high-energy region,
 $\gsim$\,TeV, once a calorimeter is set well behind 
  the vertex point of p-p interaction.
 In practice, cosmic-ray beam data 
is often obtained with an emulsion chamber (EC),
covering $E_0$\,=\,10\,$\sim$\,300\,TeV [24, 25]. 
The EC used in these references consists of approximately 
 2 m air-gap spacer between an artificial carbon target and the calorimeter,
 which is distant enough to separate each $\gamma$-ray core
 and  measure simultaneously
its shower energy using a nuclear emulsion plate.

Besides this data, we also use the recent LHCf data [26, 27] 
 on the $\gamma$-ray production energy spectrum, $dN/dE_\gamma$.
This data is taken  in the forward cone with unprecedented precision 
by placing  
  the calorimeter far away, $\approx$\,140m,
from the vertex point of p-p interaction. 
These papers give $dN/dE_\gamma$ at two energies 
in the center of mass frame (C.M.F.),
  $\sqrt{s}$\,=\,900\,GeV and 7\,TeV, corresponding to 
 L.F.\ energies of  
 approximately $E_0$ = 400\,TeV and 26\,PeV, respectively.

With the use of these data in the higher energy region as well as other past
 machine data in the lower energy region $\lsim $\,TeV, 
 we have already presented $\sigma_{pp\rightarrow \gamma}(E_0, E_\gamma)$ 
empirically in Paper I [1]. 
In this paper, we apply our work to the production cross section
 $\sigma_{pp\rightarrow e^\pm}(E_0, E_e)$ 
 of e$^\pm$ in p-p collisions, taking both the isobar
and pionization components into account, 
as discussed in Paper I. It is worth mentioning some points about
 the modeling of the nuclear interaction that were not treated in Paper I.

Historically, multiple meson production in p-p collisions was 
introduced using the fireball model [28-32]. Resonance production 
was modeled by isobar emission [33-35].
 After these early works, Stecker [15,16] proposed 
a two-component model consisting of isobar (excited baryon) 
and pionization (fireball) components, and applied this model to
 the study of cosmic $\gamma$ rays.
 Phenomenological interpretation of these 
production processes are nowadays well understood  in the framework of QCD, though we
do not use that approach here. 

The present model is based on a two-component model, 
but differs in some important respects from the model developed earlier.
For instance, we do not consider the fine structure of individual resonances
 with various masses and widths, but approximate them by 
an excited baryon with {\it effective} mass
  $M_\Delta$ and a broad width  given by ${\it \Gamma}_\Delta$,
 regarding isobaric production through a single giant resonance
 in nuclear physics. They are determined so that
the experimental data on both the 
average multiplicity and the production cross section of secondary
 products are well reproduced, as discussed in detail in Section 4.
 In Section 5.2, we compare three cross sections thus obtained, 
$\sigma_{pp\rightarrow \gamma}$, $\sigma_{pp\rightarrow e^+}$, and 
$\sigma_{pp\rightarrow e^-}$, with those obtained by 
  Dermer [19, 20], Kamae et al.\ [21] 
and PYTHIA-code [36]. 
Finally in Section 6, we discuss briefly the application of
 the present model to the calculations for the emissivity of positron, 
particularly taking nuclei effect into account.

\section{Energy spectrum of pionization secondaries}

\subsection{Modified rapidity}

Before going into the main subject where we obtain the various
 cross sections, we introduce the variable $\eta$ defined by
$$
\vspace{-2mm}
 \eta = \frac{1}{2} \ln \frac{1+\beta}{1-\beta}
  = \ln (\gamma+\sqrt{\gamma^2-1}),
 \eqno{\rm(1)}$$
with $\beta=v/c$ and $\gamma \equiv 1/\sqrt{1-\beta^2}$.
Here {\it v} is particle velocity (in the following,
we refer to $\beta$ as particle speed).

 The variable $\eta$ corresponds to the
 familiar rapidity {\it y}, which is defined by 
$$
 y = \frac{1}{2} \ln \frac{1+\beta \cos \theta}{1-\beta \cos \theta},
$$
where $\theta$ is the emission angle of the particle. As can be readily seen,
 $|\eta|$ is the
 maximum rapidity of the particle, which lies in the range $-|\eta| \le y \le |\eta|$.
We therefore, for simplicity, call $\eta$  the rapidity in the following
 discussion.

 For the numerical calculations,
 we use the second relation in Eq. (1), since
 the first expression diverges for $\beta \rightarrow 1$.
 Hence we define the modified
rapidity variable 
$$
\eta_s \equiv \eta(\gamma_s) = 
\ln (\gamma_s + \sqrt{\gamma_s^2-1}),
 \eqno{\rm(2)}
$$
regarding it 
as a function of the Lorentz factor $\gamma_s$ of particle {\it s}.
With the use of $\eta_s$ for the secondary particle {\it s} 
 with mass $m_s$, we have
 well-known expressions for the energy and momentum in natural units
 ($c =1$), namely
$$
  E_s = m_s \cosh \eta_s\;,\ \ P_s = m_s \sinh \eta_s,
$$
 and 
$$
      d \eta_s = dE_s/P_s.
$$

\subsection{Pion decay products}

We introduce a {\it normalized} energy distribution function
 of pions produced by p-p collision in the L.F., $\Phi(E_0, E_\pi)$, where $E_0$ is the
 {\it kinetic} energy of a projectile proton, and $E_\pi$ is the {\it total}
 energy of parent pions ($\equiv$\,$\pi^0,\,\pi^\pm$). 
Then the normalized energy 
distribution of decay product $\lq \lq s"$ ($\equiv \gamma,\ \mu^\pm$),  
$\varphi(E_0, \epsilon_s)$, is given by [37, 38]
$$
\vspace{-3mm}
\varphi(E_0, \epsilon_s) = 
\int_{E_\pi^-}^{E_\pi^+}
\frac{\Phi(E_0, E_\pi)}{1-\kappa_s^2} \frac{dE_\pi}{P_\pi},
 \eqno{\rm(3)}
$$
with
$$
\vspace{3mm}
\frac{E_\pi^{\pm}}{E_s}
= \frac{(1+\kappa_s^2) \pm (1-\kappa_s^2)\beta_s}{2\kappa_s^2}, 
 \eqno{\rm(4)}
$$
 where $\kappa_s = {m_s}/{m_\pi}$,  $P_\pi$ is the momentum of the parent
 pion, and $\epsilon_s$ ($E_s$) the kinetic (total) energy of  decay product {\it s}. 
$E_\pi^-$ ($E_\pi^+$) is the minimum (maximum) energy of the parent pion
 to produce the decay product $s$ with  energy $\epsilon_s$, 
 while $E_\pi^+$ = $E_0$ for $E_\pi^+ \ge E_0$.

  Note that Eq. (4) holds even in the case of
 $\kappa_s = 0$ with $[m_s, \beta_s]$ $\rightarrow$ $[0, 1]$ in
 $\pi^0\rightarrow 2\gamma$ decay. See Eq. (5) below for the explicit
 form of $E_\pi^{\pm}$ in such a decay mode.
So $\varphi(E_0, \epsilon_s)$ is 
 the important function to determine.
 
\subsection{Energy spectrum of $\gamma$ rays in
$\pi^0\rightarrow 2\gamma$ decay}

As mentioned in the Introduction, it is difficult to determine
the distribution function $\Phi(E_0, E_\pi)$ experimentally in the high energy
 region $E_0$\,$\gsim$\,1\,TeV.
 Thus we focus upon $\varphi(E_0, \epsilon_\gamma)$ instead of 
$\Phi(E_0, E_\pi)$.
In Paper I [1], we presented an empirical form based upon the
 raw machine data on $\gamma$ rays without relating it back to 
 the parent $\pi^0$. 

First we present the variables $E_\pi^\pm$
 appearing in Eqs. (3) and (4). For $``s"\equiv\gamma$, [39]
$$
E_{\pi}^-  = E_\gamma + \frac{\ m_{\pi^0}^2}{4E_\gamma},\ \  E_\pi^+ = E_0. 
 \eqno{\rm(5)}
$$
Another expression for $E_{\pi}^-$ is 
$$
E_{\pi}^- = m_{\pi^0} \cosh {\eta}_\gamma,
 \eqno{(6)}
$$
with 
$$
{\eta}_\gamma = \ln (2E_\gamma/m_{\pi^0}),
 \eqno{(7)}
$$
 which are useful for giving the upper and lower limits of
 pion energy in electron production, as discussed in the next
 section.

The production spectrum of $\gamma$ rays with
average multiplicity $\bar{N}_\gamma(E_0)$ is given by [1] 
$$
\frac{dN_\gamma}{dE_\gamma} =
\bar{N}_\gamma(E_0) \varphi(E_0, \epsilon_\gamma), 
\eqno{(8)}
$$
with
$$
\frac{\varphi(E_0, \epsilon_\gamma)}
{\beta_{c}^2{\it \Theta}_{c}} =
\frac{1}{M_{p}}
\int_{\omega_-}^{\omega_+} \hspace{-1mm}
\frac{(1-x_\gamma{\it \Gamma}_\theta)^4}
{{\it \Gamma}_\theta +\zeta \tau_\theta}
  \mbox{e}^{-\tau_\theta x_\gamma} d\omega. 
\eqno{(9)}
$$
Here $\omega= \cos \theta$, and 
$$
 x_\gamma = \frac{E_\pi^-}{E_0} = \frac{E_\gamma}{E_0}
 \biggl[1 + \frac{m_{\pi^0}^2}{4E_\gamma^2}\biggr],  
 \eqno{(10)}
$$

 with
$$
\vspace{2mm}
 \omega_{\pm} = \beta_c^{-1}
 - \beta_c^{\pm 1} \mbox{e}^{\pm \eta_0^* -\eta_{\gamma}^*},
 \eqno{(11)}
$$
$$
\eta_0^* = \ln (2\beta_c \gamma_c {M_{p}}/{m_{\pi^0}}),
 \eqno{\rm(12a)}
$$
$$
\vspace{2mm}
\eta_{\gamma}^* = \ln (2\beta_c \gamma_c {E_{\gamma}}/{m_{\pi^0}}).
 \eqno{\rm(12b)}
$$
 Here $\beta_c$ ($\gamma_c$) is the velocity
 (Lorentz factor) of the C.M.F. with respect to the L.F., 
${\it \Theta}_c$ is the normalization constant given by 
Eq. (A2), and 
$M_{p}$ the mass of proton; see Appendix A for $[\tau_\theta, 
{\it \Gamma}_\theta]$ and $[{\it \Theta}_c, \zeta]$. 
In Eq. (11), $\omega_- = -1$ for 
$E_\gamma < \frac{1}{2}m_{\pi^0}
\mbox{e}^{-(\eta_0^* + \eta_c)}$ (note that $\eta_c$ is the rapidity of the 
C.M.F.\ as measured in the L.F., and $\eta_c$ $\equiv$ $\eta(\gamma_c)$ in Eq.\ [2]).

\subsection{Energy spectra of $\mbox{e}^{\pm}$ via 
$\pi^\pm$-\,$\mu^\pm$-\,$\mbox{e}^{\pm}$ decay}

An electron (positron) with energy $\check{E}_e$ 
is created following three-body decay
of a fully polarized muon in the decay $\mu^\pm \rightarrow \mbox{e}^\pm 
+ \nu_e(\bar{\nu}_e) + \bar{\nu}_\mu({\nu}_\mu)$.
The production of $\mu^\pm$ is kinematically equivalent to 
the production of $\gamma$ rays because the two-body decay is isotropic, as
 discussed in Section 2.2.

 The energy/angular distribution of electrons (positrons) 
in the muon rest frame is given by [51-53]
\vspace{-1mm}
$$
\vspace{1mm}
\check{f}(\check{q}, \check{\theta}) =\check{q}^2 (3-2\check{q})
\biggl[1 + \xi \cos \check{\theta}
\frac{1-2\check{q}}{3-2\check{q}}\biggr],
\eqno{(13)}
$$
where $\check{q}=2\check{E}_{e}/m_\mu$,
  $\xi=+1$ ($-1$) for electrons (positrons),
 and $\check{\theta}$ is the angle between the electron (positron) 
and the spin of the muon in the muon rest frame.
 Eq. (13) is valid also for muon neutrinos $\nu_\mu$ 
in the approximation that the electron mass $m_e$ is neglected.
For the electron neutrino $\nu_e$, 
 both ($3-2\check{q}$) and ($1-2\check{q}$) in Eq. (13) 
are replaced by $6(1-\check{q})$ [53]. 

In practice we need the energy distribution of $\mbox{e}^\pm$,
  $f(E_\pi, E_e)$, for a parent pion with the {\it fixed} 
energy $E_\pi$ in the L.F.. The full expression for $f(E_\pi, E_e)$, which 
is tedious to evaluate, is summarized  
 by Dermer ([19]; see also Moskalenko \& Strong [54]).

For the pion rapidity $\eta_\pi$,
we introduce a function $\phi(\eta_\pi, q_e)$ defined by 
\vspace{-2mm}
$$
 \phi(\eta_\pi, q_e) = {P_\pi}f(E_\pi, E_e), 
\eqno{\rm (14)}
$$
and in Table 1 we summarize it explicitly with the use of three parameters,
$[q_e, \eta_\pi, \tilde{\eta}_\mu]$, 
which are given  by
\vspace{-3mm}
$$
q_e = 2 E_e/m_\mu,
\eqno{\rm (15a)}
$$
and
$$
\vspace{3mm}
E_\pi = m_\pi \cosh \eta_\pi,
\eqno{\rm (15b)}
$$
with $
\eta_\pi \equiv \eta(\gamma_\pi)$, and 
$\tilde{\eta}_\mu \equiv \eta(\tilde{\gamma}_\mu)$ in Eq. (2).
The term $q_e$ is the L.F.\ electron (positron) energy $E_e$, 
in units of half of the muon mass, $\frac{1}{2}m_\mu$. Two rapidity variables, 
$\eta_\pi$ and  $\tilde{\eta}_\mu$,  correspond to 
those of the parent pion in the L.F.\  and
  the daughter muon  in the pion rest frame, respectively. 
Since $\tilde{\gamma}_\mu = \frac{1}{2}(m_\mu/m_\pi + m_\pi/m_\mu) = 1.04$,
 we find $\tilde{\eta}_\mu = 0.278$ with $\tilde{\beta}_\mu = 0.271$, 
 and $[g_0;\ g_{1,0}$, $g_{2,0}]$ in Table 1 equal [1.16; 3.75, 6.78] for 
$\xi$ = 1 (electron), and [-0.57; 4.49, 1.03] for $\xi$ = -1 (positron), 
respectively.

In Fig. 1a, we present numerical values of $\phi(\eta_\pi, q_e)$
 vs.\ $q_e\exp[-(\eta_\pi + \bar{\eta}_\mu)]$ for three
 pion kinetic energies $\epsilon_\pi$ ($\equiv$ $E_\pi - m_\pi$)
 = 0.01, 0.1, and 1\,GeV in the case of electrons (green curves) and 
positrons (red curves). For $\epsilon_\pi$ $\gsim$ 1\,GeV,
one finds $\phi(\eta_\pi, q_e)$ $\approx$
$\phi(q_e\mbox{e}^{-\eta_\pi})$. Hence in the high energy region, 
$q_e\mbox{e}^{-\eta_\pi} \approx E_e/E_\pi$, so that 
$\phi(\eta_\pi, q_e) \approx \phi(E_e/E_\pi)$.

The normalization in Eq. (14) holds exactly,
 taking care that the restriction $q_e \subseteq [q_e^-, q_e^+]$
 in Table 1, namely
$$
\int_{E_{e}^{-}}^{E_{e}^{+}}
 \phi(\eta_\pi, q_e)\frac{dE_e}{P_\pi} = 1,
$$
where
$$ 
\vspace{1mm}
E_{e}^{-}=0;\ \ 
2E_{e}^{+}
 = m_\mu\mbox{e}^{\eta_\pi + \tilde{\eta}_\mu},
$$
is satisfied.

Once we have the normalized energy spectrum of $\mbox{e}^\pm$ 
coming from  the $\pi^\pm$-$\mu^\pm$-${\rm e}^\pm$ decay in the L.F., 
we  can straightforwardly obtain the production 
 energy spectrum of $\mbox{e}^\pm$ in p-p collisions
 (see Eq.\ [3] for $\Phi$), through the expression
$$
\frac{dN_{e^\pm}}{dE_e} = \bar{N}_{\pi^\pm} \! \!
\int_{E_\pi^-}^{E_\pi^+} \! \Phi(E_0, E_\pi)
f(E_\pi, E_e)dE_\pi,
\eqno{\rm (16)}
$$
where
$$
\vspace{1mm}
E_\pi^\pm = m_\pi \cosh \eta_\pi^\pm.
\eqno{\rm (17)}
$$
Here
$\bar{N}_{\pi^\pm}(E_0)$ is
 the effective multiplicity of
 charged pions, which depends on the projectile proton energy $E_0$,
and
$$
\eta_\pi^- = \bar{\eta}_e - \tilde{\eta}_\mu;\ \ 
\eta_\pi^+ = \bar{\eta}_\pi^* + \eta_c.
\eqno{\rm (18)}
$$
Here $\bar{\eta}_e = \ln q_e$, and 
  $\bar{\eta}_\pi^*$ is the maximum C.M.F.\ rapidity of pions
given by Eq. (A3) in Appendix A, and $\eta_\pi^+$ is the maximum pion
rapidity in the L.F..

Another expression for $E_\pi^-$ in Eq. (17) is given by
$$
E_\pi^- = E_e + \frac{m_\pi^2}{4E_e},
\eqno{\rm (19)}
$$
which is similar to Eq. (5) for the $\gamma$-ray
energy spectrum.

Noting the relation between $\varphi$ and $\Phi$ given by Eq. (3)
 with $E_\pi^-(\epsilon_\gamma)$
 in the case of $\pi^0 \rightarrow 2\gamma$ decay, 
 we have\,\footnote{Note an additional constant term in Eq. (3),
 $(1-\kappa_s^2)^{-1}$ ($\kappa_s$\,$\neq$\,0 in $\pi$-$\mu$\,decay),
 which cancels the normalization term 
$(1-\kappa_s^2){\it \Theta_c}$.}
$$
\vspace{3mm}
\frac{\partial \varphi}{\partial \epsilon}
 \biggr |_{\epsilon = E_\pi - m_\pi}
    = - \Phi (E_0, E_\pi) \frac{1}{P_\pi},
\eqno{\rm (20)}
$$
and using Eq. (14) together with the relation $dE_\pi/P_\pi = d \eta_\pi$,
 eventually we obtain
$$
 \frac{dN_{e^\pm}}{dE_e} = {\bar{N}_{\pi^\pm}}
 \! \int_{\eta_\pi^-}^{\eta_{\pi}^+} \! \varphi(E_0, \epsilon_\pi)
\phi^\dagger (\eta_\pi, q_e)d\eta_\pi,
\eqno{\rm (21)}
$$
where $\phi^\dagger (\eta_\pi, q_e)$ denotes the differential 
of $\phi(\eta_\pi, q_e)$ with respect to $\eta_\pi$. An explicit form of
 $\phi^\dagger  (\eta_\pi, q_e)$ is presented in Table 1
 together with $\phi(\eta_\pi, q_e)$.
See Eq. (18) for $\eta_\pi^{\pm}$, recalling that 
$\epsilon_\pi$ ($\equiv E_\pi-m_\pi$) is the L.F.\ kinetic energy of 
 pion expressed in terms of $\eta_\pi$ as
$$
\epsilon_\pi = 2 m_\pi \sinh^2 \frac{\eta_\pi}{2} .
$$

Fig. 1b shows numerical values of $\phi^\dagger(\eta_\pi, q_e)$
 vs.\ $q_e\exp[-(\eta_\pi + \bar{\eta}_\mu)]$, for the same
 three parent pion kinetic energies used in
  Fig. 1a. We find again a scaling behavior of 
$\phi^\dagger(E_e/E_\pi)$ similar to  $\phi(E_e/E_\pi)$
 in the high energy region $\epsilon_\pi\ \gsim\ 1$\,GeV.

\section{Energy spectrum of secondary products
 in isobaric resonances}

In the low energy region around $E_0\cong 1$ -- 3 GeV in p-p collisions, 
many isobaric resonances, such as
${\it \Delta}(1232)$, ${\it \Delta}(1410), \ldots $\,, 
 as well as the exclusive channel $pp \rightarrow \pi^+ d$
 ($d$: deuteron)  become
 effective for secondary production in addition to those coming from
 the pionization component mentioned in the last section.
 In this section we consider an expression for
the production cross section of secondaries that approximately
reproduces the experimental data, particularly for
 $\pi^0$ and $\pi^+$. This does not necessarily give a good 
representation for other mesons, such as
$\pi^-$ and K$^{\pm}$ (see Figs.\ 2a and 2b), but the contributions of
 which are negligibly small comparing to those of  $\pi^0$ and $\pi^+$.

\subsection{Energy spectrum of $\gamma$ rays via $\Delta$\,-\,$\pi^0$-
             $2\gamma$ decay}

We assume that an isobar with mass $M_\Delta$ is produced in p-p
 collision, and disintegrates  isotropically into $\pi^0$ (+ $p$) or
$\pi^+$ (+ $n$) with pion energy 
$\hat{E}_{\pi}$ in the {\it isobar} rest frame.
This implies
$$
\hat{E}_{\pi} = m_\pi \tau_{\Delta} \cosh \hat{\eta}_\Delta,
\eqno{\rm (22)}
$$
with
$$
\hat{\eta}_\Delta = \ln (\tau_\Delta M_\Delta /m_\pi);\ \ 
\tau_{\Delta} = \sqrt{1 - M_{p}^2/M_{\Delta}^2}.
$$

In the calculations,  we set $M_{\Delta}$\,=\,1.25\,GeV/c$^2$ with 
$\tau_{\Delta} =0.66$ (see Sec.\ [4.1]),
so that  numerical values of the decaying pion in the isobar rest 
frame, $\hat{E}_{\pi}$, $\hat{P}_{\pi}$\,(momentum), 
$\hat{\beta}_\pi$\,(velocity),
$\hat{\gamma}_\pi$\,(Lorentz factor), and $\hat{\eta}_\pi$\,(rapidity),
 are given by 
$$
[\hat{E}_{\pi},\ c\hat{P}_{\pi}] = 
[279,\ 242]~ \mbox{MeV},
$$
and
$$
[\hat{\beta}_\pi, \hat{\gamma}_\pi, \hat{\eta}_\pi] = [0.87, 2.01, 1.32],
$$
 respectively.

One might argue that there are many resonances disintegrating
 into $\pi^0$ or $\pi^\pm$, in contrast with a single resonance with
 $M_\Delta$\,=\,1.25,GeV/c$^2$ assumed here. As emphasized in the 
Introduction, however,
 we do not consider the {\it fine} structure or mass spectrum of individual
 resonances, 
but approximate resonance production in terms of something 
 like a giant resonance (well-known in nuclear physics), the width of which is
 determined so that the experimental data are  consistently reproduced 
 with Eq. (29). 
Moreover, even in more detailed treatments that employ a Breit-Wigner distribution 
for the resonance mass spectrum, there is an implicit assumption that the
 excited 
nucleon travels in its original, pre-collision direction. This may be a good 
assumption for low-momentum transfer, peripheral collisions.

Therefore, the energy distribution of pions via isobar disintegration 
in the L.F.\ can be written as
$$
\vspace{2mm}
\Phi_\Delta (E_0, E_\pi) dE_\pi = \Phi_0(E_0) dE_\pi,
\eqno{\rm (23)}
$$
and $\Phi_0$ is determined from the normalization bound to 
 $\hat{E}_\pi^- \le E_\pi \le \hat{E}_\pi^+$ with
$$
\hat{E}_\pi^\pm = \gamma_\Delta^* \gamma_c[(1+\beta_c\beta_\Delta^*)
\pm (\beta_c+\beta_\Delta^*)\hat{\beta}_\pi]\hat{E}_\pi,
$$
where $\gamma_\Delta^*$ ($\beta_\Delta^*$) is Lorentz factor (velocity) of the
 isobar in C.M.F. given by 
$$
 \gamma_\Delta^* = \tau_\Delta \cosh \eta_s^*;\ \ 
\eta_s^* = \ln (\sqrt{s}/\tau_\Delta M_\Delta),
$$
and $\sqrt{s}=2M_{p}\gamma_c$ is the center of mass energy.

Using  isobar rapidities $\eta_\Delta$ and $\eta_\Delta^*$,
in the L.F.\ and C.M.F.\ respectively, we have
 a simple and physically trivial expression,
$$
\hat{E}_\pi^\pm = m_\pi \cosh (\eta_\Delta \pm \hat{\eta}_\pi),
\eqno{\rm (24)}
$$
with
$$
\eta_\Delta = \eta_\Delta^* + \eta_c \equiv  \eta(\gamma_\Delta^*)
   + \eta_c.
$$

Now, the normalization constant $\Phi_0(E_0)$ in Eq. (23) is given by
$$
\Phi_0(E_0) 
=\frac{1}{2\hat{P}_{\pi}}\frac{1}{\sinh \eta_\Delta}.
\eqno{\rm (25)}
$$
and from Eq. (3), we can write 
 the production energy spectrum of $\gamma$ rays coming from the isobaric 
 resonances with $[E_\pi^-, E_\pi^+]$ given by Eq. (5), namely
$$
\varphi_\Delta(E_0, \epsilon_\gamma) = 
\int_{E_\pi^-}^{E_\pi^+}\! \! 
{\Phi_\Delta(E_0, E_\pi)} \frac{dE_\pi}{P_\pi}.
$$

Using the relation $d\eta = dE_\pi/P_\pi$, finally we have a simple form
$$
\frac{dN_\gamma^\Delta}{dE_\gamma} =
\frac{\bar{N}_\gamma^{\Delta}}{2\hat{P}_{\pi} \sinh \eta_\Delta}
 [\eta_{p} - \eta_\gamma],
\eqno{(26)}
$$
where ${\bar{N}_\gamma^{\Delta}}$ is the effective multiplicity
 of $\gamma$-rays (via $\pi^0$) through the $\Delta$-isobar
 disintegration, and 
$\eta_{p}$ is the rapidity of projectile proton
 given by Eq. (2) with $\eta_p \equiv \eta(\gamma_p)$, 
with $\gamma_p = 1 + E_0/M_p$, 
 and $\eta_\gamma$ the minimum rapidity of pion given by
 Eq. (7) to produce $\gamma$ rays with  energy 
$E_\gamma$\,(=\,$\epsilon_\gamma$). 

\subsection{Energy spectrum of e$^+$ via 
$\Delta$\,-\,$\pi^+$-\,$\mu^+$- $\mbox{e}^+$ decay}

As the contribution of $\pi^-$ from the isobaric resonance in p-p 
 collision is
 negligible in comparison with $\pi^+$ around a few GeV
(see Fig.\ 2b), 
we consider here the energy spectrum of e$^+$ via 
$\pi^+$-\,$\mu^+$- $\mbox{e}^+$ decay only. 

Similar to Eq. (16), we obtain
 the production energy spectrum of e$^+$ coming from the isobar decay, 
$\Delta$\,$\rightarrow$\,$\pi^+$$\rightarrow$\,$\mu^+$\,$\rightarrow$\,$\mbox{e}^+$, 
namely
$$
\frac{dN_{e^+}^\Delta}{dE_e} = \bar{N}_{\pi^+}^\Delta \! \!
\int_{E_\pi^-}^{E_\pi^+} \! \Phi_\Delta(E_0, E_\pi)
f(E_\pi, E_e)dE_\pi,
$$
where $\bar{N}_{\pi^+}^\Delta$ is the effective multiplicity of
 $\pi^+$ through the disintegration of the $\Delta$-isobar, noting
 Eq. (17) for $E_\pi^\pm$.

With the use of $\Phi_0(E_0)$ given by Eq. (25) 
in $\Phi_\Delta(E_0, E_\pi)$, we
 obtain
$$
\vspace{2mm}
\frac{dN_{e^+}^\Delta}{dE_e} =
\frac{\bar{N}_{\pi^+}^\Delta}{2\hat{P}_{\pi} \sinh \eta_\Delta} \! 
\int_{\eta_{\pi}^-}^{\eta_{\pi}^+}
\! \!{\phi(\eta_\pi, q_e)d\eta_\pi},
\eqno{\rm (27)}
$$
where $\eta_\pi^\pm$ are given by Eqs. (18) and (A3). 

Here it is important to take care that the
 integration with respect to $\eta_\pi$ in the above equation, which must be
 performed separately over the allowed range of $\eta_\pi$,  depending 
  on positron energy $E_e\ (=\frac{1}{2}m_\mu q_e)$; 
 see  the restriction $q_e \subseteq [q_e^-, q_e^+]$ in Table 1.
Note also that the energy spectra of  secondary 
 $\gamma$ rays and $\mbox{e}^+$ coming from both pionization and 
isobaric components given by Eqs. (8), (21), (26) and (27), 
are all expressed only by $\phi$,  $\phi^\dagger$, and 
$\int\! \phi\,d\eta_\pi$, apart from the basic function
$\varphi(E_0, \epsilon)$ for the pionization components in
 Eqs. (8) and (21).
 In the next section, explicit numerical values of the parameters 
 appearing above, $M_\Delta,\ \bar{N}_\gamma^\Delta$, etc., 
are given.

\section{Multiplicity of secondary particles}

We presented in the previous section the production energy spectra  
 of $\gamma$ rays and $\mbox{e}^\pm$ in p-p collisions,
[$dN_{\gamma}/{dE_\gamma}$, $dN_{e^\pm}/{dE_e}$] from the
 pionization components,
 and [$dN_\gamma^{\Delta}/{dE_\gamma}$, $dN_{e^+}^\Delta/{dE_e}$] 
 from the isobaric ones, respectively. 
For these spectra, we need  the {\it average} multiplicity of secondaries
 per p-p collision, 
[$\bar{N}_\gamma$, $\bar{N}_{\pi^\pm}$] for the former components, 
 and [$\bar{N}_\gamma^\Delta$, $\bar{N}_{\pi^+}^\Delta$] for the
 latter ones, respectively, each 
depending on the energy of projectile proton $E_0$.
 These numerical values should be determined from the machine data.

Here, as noted also in Paper I,
the multiplicity 
 should be regarded as an {\it effective} one rather than an {\it actual} one,
 particularly in the high energy region, since the low-energy secondaries
 produced in the backward cone in the C.M.F.\ in p-p collision do not make
a signficant contribution to the total secondaries formed 
in the galactic CR system.

\subsection{Parameterization of the empirical formulae}

 In this paper, we  modify slightly the average multiplicity of $\gamma$'s,  
$\bar{N}_\gamma (E_0)$, presented in Paper I, since those coming
 from isobar decay, such as $\bar{N}_\gamma^\Delta (E_0)$ from $\pi^0$\,+\,p given by Eq. (29) below,  
 are additionally included in the present work. 
Hence the {\it total} average multiplicity,
 $\bar{N}_\gamma$\,+\,$\bar{N}_\gamma^\Delta$, 
 must be fitted to the machine data in contrast to the fitting
 without $\bar{N}_\gamma^\Delta$ in Paper I. 

We assume that the energy dependence on the average multiplicities 
of $\pi^\pm$ and K$^\pm$ are of the same form as in the case of $\gamma$ rays,
 for both pionization and isobar components, 
$\bar{N}_s (E_0)$ and $\bar{N}_s^\Delta (E_0)$ 
($\lq\lq s"\,\equiv\,\gamma, \pi^\pm, \mbox{K}^\pm$), respectively,
 but with different numerical values in the parameterization 
 for each term as discussed below.

First, we summarize the multiplicity coming from the pionization 
 component, slightly modifying the parameterization in Paper I, now in 
the form
$$
\bar{N}_s(E_0) = \bar{N}_{0} 
{\it \Lambda}_s({\epsilon}_0) {\epsilon}_0^{\,0.115},
\eqno{\rm (28)}
$$
with
$$
{\epsilon}_0 = E_0 - E_{\rm th},
$$
and
$$
{\it \Lambda}_s(\epsilon_0) =
\Bigl[1- \mbox{e}^{-\sqrt{\epsilon_0/\epsilon_1}}\Bigr]
\Bigl[1- \mbox{e}^{-\hspace{-0.5mm}\sqrt[4]{\epsilon_0/\epsilon_2}}\Bigr]
\mbox{e}^{-(\epsilon_c/\epsilon_0)^k},
$$
where numerical values of $\bar{N}_0$, $E_{\rm th}$, 
 $\epsilon_c$, and $k$ are
 presented in Table 2 for individual secondaries, $\lq \lq s$" $\equiv$ 
 $\gamma, ~\pi^\pm, \mbox{K}^\pm$, while $[\epsilon_1, \epsilon_2]$ 
 = [4.53\,GeV, 1.98\,TeV] irrespective of {\it s}.

Second, for the multiplicity of $\gamma$ rays and $\pi^+$ through the 
disintegration of a $\Delta$ isobar, we assume 
$$
\vspace{1mm}
\bar{N}_s^\Delta (E_0) = \bar{N}_{0}^\Delta
\mbox{exp}\biggl[-
\biggl\{\frac{\ln ({\epsilon}_0/M_\Delta)}{\ln m_\pi}\biggr\}^2
\biggr],
\eqno{\rm (29)}
$$
where $m_\pi$ is the pion mass in GeV/c$^2$, and 
$\bar{N}_0^\Delta$ = [0.61, 0.25]  for 
$\lq \lq s" \equiv$ [$\pi^+, \gamma$] respectively.
 One can regard the exponential function in Eq. (29)
 as the width in $\ln M_\Delta$-scale of the $\Delta$ isobar with the mass
 $M_\Delta$ instead of the Breit-Wigner type function.

We assume $M_\Delta$ = 1.25 GeV/c$^2$,
 leading to [$\hat{E}_\pi,
\hat{\eta}_\pi$] = [0.279\,GeV, 1.32] as presented in Section 3.1.
These numerical values are determined so that the experimental data 
on the total production cross sections of $\gamma$ rays,
 $\pi^\pm$,
 and K$^\pm$, as presented in Figs. 2a and 2b, are reproduced.

\subsection{Comparison with total production cross section data}

 The total production cross section, $\sigma_{pp \rightarrow s}^{tot}$, 
 to produce a secondary $\lq\lq s$"
coming from both pionization and isobar components is given by 
$\sigma_{inel} \times (\bar{N}_s + \bar{N}_s^\Delta)$, where
$\sigma_{inel}$ is the total inelastic collision cross section,
including all collisions except elastic one.
In the following, we simply rewrite $\bar{N}_s$\,+\,$\bar{N}_s^\Delta$ with 
 $\bar{N}_s$, unless mentioned specifically,
 and regard,  for simplicity, $\bar{N}_s$ as the total average multiplicity of
 the element {\it s}. The explicit form of $\sigma_{inel}$ is given by 
 Eqs. (1) and (2) in Paper I, covering from the threshold energy of pion
 production to the LHC energy.

In Figs. 2a and 2b we show historical data ([1, 16, 19, 55, 56])
on  $\sigma_{pp \rightarrow s}^{tot}$
  for $\gamma$ rays (a) and charged mesons (b). Here, broken curves
 correspond to those coming from isobar component alone,
$\sigma_{inel} \times \bar{N}_s^\Delta$, see Eq. (29) for 
$\bar{N}_s^\Delta$, and the
 solid ones from the summed cross sections, 
$\sigma_{inel} \times (\bar{N}_s$\,+\,$\bar{N}_s^\Delta)$, see
Eq. (28) with Table 2 for  $\bar{N}_s$.
Additionally  plotted by blue circles 
 is the very high energy LHC data with $\sqrt{s}$ =
 900\,GeV, 7\,TeV (Paper\,I [1]). 
 In Fig.\ 4b of Appendix A, we demonstrate an example of this model by 
fitting   recent data at $\sqrt{s}$ = 900\,GeV [57].
 See Paper\,I for a more
complete comparison, including pseudo-rapidity, 
Feynman variable, $\gamma$-ray energy, and so on, 
 in both the C.M.F. and L.F.,
 covering very wide energies up to  LHC energies, ranging from
$E_0$ = $1\,\mbox{GeV}$ to $ \approx 20\,\mbox{PeV}$.

 Fig. 2b shows that the contribution of K$^\pm$ is on average
 as large as 7\% by total number of charged mesons made when $E_0$ $\gsim$
 10\,GeV, and negligible below, 
so that the contribution to e$^\pm$
 coming from K$^\pm$ could be at most 7\%, and smaller after decay
 kinematics are taken into account.
Thus, in the 
 numerical calculations, 
we consider only $\pi^\pm$'s as the source of secondary e$^\pm$'s,
 but assume 7\% increase in the absolute intensities of secondary e$^\pm$'s 
 we are interested in due to kaon contribution.

\section{Comparison with other numerical codes for the elementary processes}

Many quite elaborate codes 
for the treatment of secondary nuclear production processes, 
with applications to CRs and
 galactic phenomena, have been developed ([19, 20, 21, 36, 54]).
Accurate cross sections and production spectra become
 increasingly  important as  new observational data from {\it Fermi}
 and AMS-02
 become available with high quality in both statistics and
 systematics. 
In this section, we compare our calculations for the production 
cross sections of $\gamma$ rays and $\mbox{e}^\pm$ in p-p collision 
 with those calculated by (a) Dermer [19], 
(b) Kamae et al.\ [21], and the (c) PYTHIA-code
 (Sjostrand [36]).

In order to compare their calculations with the present ones, 
 we summarize the notations for the production cross section
 of individual elements in the followings,
$$
\frac{d\sigma}{dE_\gamma}(E_0, E_\gamma) =
\sigma_{inel} \biggl(\frac{dN_\gamma}{dE_\gamma} + 
\frac{dN_\gamma^\Delta}{dE_\gamma} \biggr),
\eqno{\rm (30a)}
$$
for $\gamma$ rays, 
$$
\frac{d\sigma}{dE_e}(E_0, E_e) =
\sigma_{inel} \biggl(\frac{dN_{e^+}}{dE_{e}} +
\frac{dN_{e^+}^\Delta}{dE_{e}}\biggr),
\eqno{\rm (30b)}
$$
for positrons,  and
$$
\vspace{2mm}
\frac{d\sigma}{dE_e}(E_0, E_e) =
\sigma_{inel} \frac{dN_{e^-}}{dE_{e}},
\eqno{\rm (30c)}
$$
for electrons. See Paper I for $\sigma_{inel}$. 

Explicit equations used for the calculations of the spectra in the
 right hand side correspond to following relations: 
$$
\biggl[\frac{dN_{\gamma}}{dE_\gamma},\ \frac{dN_\gamma^{\Delta}}{dE_\gamma}
\biggr]
 \Rightarrow [{\rm Eq.\ 8},\ {\rm Eq.\ 26}],
\eqno{\rm (31a)}
$$
$$
\vspace{2mm}
\biggl[\frac{dN_{e^\pm}}{dE_e},\ \frac{dN_{e^+}^\Delta}{dE_e}\biggr]
 \Rightarrow [{\rm Eq.\ 21},\ {\rm Eq.\ 27}].
\eqno{\rm (31b)}
$$

\subsection{Production cross section of $\gamma$ rays}

Figs. 3(a1-c1) show our numerical results on 
the production cross section
of $\gamma$ rays in p-p collision (heavy solid curves) in 
comparison with
numerical results by (a) Dermer [19] and Murphy et al.\ [41], 
(b) Kamae et al.\ [21] and
Karlsson \& Kamae [41],
 and the (c) PYTHIA-code [36], for $E_0$ = $10^0, 10^1, 10^2, 10^3, 10^4$\,GeV.
 Note that Murphy et al.\ presents results only up to $E_0$ = $10^2$ GeV, and
 PYTHIA is not applicable at low, $E_0\,\lsim$\,1\,GeV, energies. 

As can be seen, the models are in general agreement.
Below 1 GeV, however, our model with the simplified treatment
of resonance production deviates from the Kamae et al.\ model, but
is in good agreement with the Murphy et al.\ calculations.
 Note also that Kamae model is not symmetric about $\frac{1}{2}m_{\pi^0} \approx 70$\,MeV
 when plotted vs $\ln E_\gamma$, while it must be kinematically symmetric
 as is well-known. 

 Small but significant deviations are also seen in the production spectra when
$E_0\,\gsim\,1\,$TeV, increasing with energy.
These differences, which would earlier be masked by the large observational uncertainties, are now important when treating
 secondary particles produced by CR/ISM collisions.
Machine data used in past models focused on the energy range
 $E_0\,\lsim\,$TeV, while this range is now becoming
important due to the improved high-energy CR data.
Fig. 4b in 
Appendix A is an example of the comparison between the present
 parameterization and the most recent LHCf data [57] on the production 
energy spectrum of $\gamma$ rays.\footnote{See Paper\,I
  for the additional comparisons, not only for the energy
 spectrum, but also for  the pseudo-rapidity
 distribution of $\gamma$ rays at TeV energies and higher.}

 Ground-based very-high-energy $\gamma$-ray astronomy,  
overlapping with  {\it Fermi}-LAT
in the energy range between $\approx 50$ -- 200\,GeV, 
will provide additional relevant data.
For example,  H.E.S.S., besides discovering 
many new TeV sources, 
has also recently announced the detection of diffuse $\gamma$ rays in 
the TeV region around the galactic plane [42]. Furthermore,   
the full observation program by the Cherenkov Telescope Array (CTA) will start around 2020 [43], 
 so that the LHC data at TeV energies and higher
 becomes increasingly relevant to analysis of TeV $\gamma$-ray data.

\subsection{Production cross section of $\mbox{e}^\pm$}

Figs. 3(a2-c2) and 3(a3-c3) show production cross sections
 of positrons and electrons in p-p collisions, respectively.
Our results are similar to those by Dermer [19]  and Murphy et al.\ [41] 
in the low-energy regime for e$^+$ production, though not 
 for electrons at $E_0=1$~GeV in Fig. 3(a3), which however make a very minor
contribution to lepton production. We again find 
 that there exists a significant discrepancy between ours and 
other models in the high-energy, $E_0 \gg 100$ GeV, regime.

 Unfortunately,  we have no user-friendly experimental data on 
 the production of secondary electrons (positrons) via $\pi$-$\mu$ decay.
However, as emphasized in Section 2, the cross sections for 
$\gamma$ and $\mbox{e}^\pm$ production
 are linked through the $\pi^\pm$-decay kinematics. 
So  the reliability
 of the $\sigma_{pp\rightarrow e^\pm}(E_0, E_e)$ cross section 
depends upon the reliability
of the  
$\sigma_{pp\rightarrow \gamma}(E_0, E_\gamma)$ cross section, where data
 are available from current machine experiments.
Through this procedure, the cross sections for $\mbox{e}^\pm$ production
are accurate  over a wide energy range,
 as confirmed by our parameterization of
$\sigma_{pp\rightarrow \gamma}(E_0, E_\gamma)$, which is valid 
  even at LHC energies.

\section{Discussion}

We present that the production energy spectra of
$\gamma$ rays and $\mbox{e}^{\pm}$ in p-p collision 
are both expressed in terms of
 the common function $\varphi(E_0, \epsilon)$ (Eq.\ [9]), 
with $\epsilon$\,$\equiv$\,$\epsilon_\gamma,
 \epsilon_\pi$, which is determined by the machine data on
 $\gamma$ rays in p-p collision. 
The two production energy spectra, Eqs. (8) and (21), 
  are kinematically equivalent in the sense that 
they are linked by a {\it model-independent} function 
$\phi(\eta_\pi, q_e)$ (Table 1) 
 without referring back to $\Phi(E_0, E_\pi)$,
 the production cross section of parent pions.
 Note that the present production cross section for
e$^\pm$ is applicable also for muon- and electron-neutrinos as discussed in
Sec.\ 2.4., assumung $m_e \approx 0$, which will be studied elsewhere
 in the nearfuture.

 Since we have confirmed already in Paper I that the common function 
$\varphi(E_0, \epsilon)$ reproduces well the accelerator data over
 the wide energy range from GeV to 20\,PeV in projectile proton, 
$\sigma_{pp\rightarrow e^\pm}(E_0, E_e)$ is applicable enough
 even for PeV electron. One sholud note that Eq.\ (21) holds 
irrespective of the form of $\varphi(E_0, \epsilon)$.

Now, having focused only on p-p collisions so far, we have to
take the nuclei effects into account in practice for the application
to the study of the galactic phenomena.
In order to quantify them, the
$\lq\lq$nuclear enhancement factor" is used, which is defined by
$$
\varepsilon_{\rm H} (\vct{r}; E_s) = 
q_{{\rm all}\rightarrow s}(\vct{r}; E_s)/q_{{pp}\rightarrow s}
(\vct{r};, E_s),
$$
with $\lq \lq s$" $\equiv$ $\gamma,\ \mbox{e}^\pm,\ \bar{\rm p}$, and so on. 
Here, 
$q_{{\rm all}\rightarrow s}$ is the {\it total emissivity} of secondary $s$ 
including all CR elements (projectiles; p, He, $\ldots$\,Fe) as well as 
the helium gas contamination (targets; H, He) in the ISM,
while $q_{{pp}\rightarrow s}$ from only the p-p collision.
 Several authors give $\varepsilon_{\rm H}$
 $\sim$\,1.5 (Cavallo \& Gould [44]), $\sim$\,1.6 (Stephens \& Badhwa [18]),
  $\sim$\,1.45 (Dermer [19]), and $\sim$\,1.52 (Gaisser \& Schafer [45]),
 $\sim$\,1.53 (Shibata et al.\ [46]).

As discussed in Paper I, $\varepsilon_{\rm H}$ does not depend so strongly
 on the nuclear interaction model, but on the composition of
 both projectile (CR's) and the target nuclei (ISM).
In fact, Mori [47] recently takes account of heavy nuclei other than 
helium in the ISM, and obtains somewhat larger values, finding
  $\varepsilon_{\rm H}$ = 1.8\,-2.0. 
 We will present $\varepsilon_{\rm H}$ elsewhere, based on the most recent
 data on
 the CR composition and spectra, with higher staistics and unprecedented
 precision by ATIC [48], TRACER [49], and CREAM [50] for heavy elements 
in addition to PAMELA [3] and ATMS02 [4] for proton and helium.

Finally, we will study the origin of positron excess nowdays established
 by PAMELA and AMS-02 in the near future, in connection with the darkmatter
 scenario [58-61], combining the present work as a
 background positron spectrum.

\begin{center}
Acknowledgments
\end{center}

We greatly appreciate C.~D.~Dermer for his careful reading of the
 present paper and valuable comments.
Two of authors (T.~S. and R.~Y) would like to express their deep appreciation for the Research Institute,
 Aoyama-Gakuin University, for supporting our research.
This work is also supported in part by Grant-in-Aid for Scientific research
 from the Ministry of Education, Science, Sports, and Culture (MEXT), Japan,
 No. 21111006, No. 22244030, No. 23540327 (K.~K.), No. 24.8344 (Y.~O.),
and supported by the Center for the Promotion of Integrated Science (CPIS)
 of Sokendai (1HB5804100) (K.~K.).

\begin{center}
APPENDIX A \\ 
Renormalization of the production cross section and 
comparison with machine data
\end{center}

First, we present the 
 renormalized constant ${\it \Theta}_c$ in contrast to the previous one 
in Paper I.
Note that
 $x = E_\gamma/E_0$ in Paper I is replaced by $x_\gamma$ in Eq. (9);
$$
 x = \frac{E_\gamma}{E_0}\ \Rightarrow \ 
  x_\gamma  = 
 \biggl[1 + \frac{m_{\pi^0}^2}{4E_\gamma^2}\biggr]x, 
$$
 which comes from the kinematical limit in $\pi^0$\,$\rightarrow$\,$2\gamma$ 
given by Eq. (5), while we used an approximation 
$E_\gamma \gg m_{\pi^0}/2$ in Paper I. So the approximation affects slightly 
 the normalization constant ${\it \Theta}_c$, but the shape of the energy
 distribution in $\sigma_{pp \rightarrow \gamma}(E_0, E_\gamma)$ is not deformed except the low energy region, $E_\gamma\, \lsim\, m_{\pi^0}/2$.
[$\tau_\theta, {\it \Gamma}_\theta, \zeta$] appearing in Eq. (9) are 
given by 
$$
\tau_\theta = 2 (\gamma_{{c}}^2 -1)
(M_p/p_0) \sin \theta,
\eqno{(\rm A1a)}
$$
$$
\vspace{2mm}
{\it \Gamma}_\theta = 
2(\gamma_{c}^2 -1) 
(1- \beta_{c} \cos \theta), 
\eqno{(\rm A1b)}
$$
 and $\zeta$ = 0.02.

 The normalization constant ${\it \Theta}_c$ is given by
$$
\frac{1}{\it \Theta_c} = \frac{\beta_c^2}{M_{\rm p}}
\int_{E_\gamma^-}^{E_\gamma^+}\! \! dE_\gamma
\int_{\omega_-}^{\omega_+} \hspace{-1mm}\frac{(1-x_\gamma 
{\it \Gamma}_\theta)^4}
{{\it \Gamma}_\theta +\zeta \tau_\theta}
  \mbox{e}^{-\tau_\theta x_\gamma} d\omega, 
\eqno{(\rm A2)}
$$
with
$$
\vspace{2mm}
  2E_\gamma^{\pm} = m_{\pi^0}\mbox{e}^{\pm (\bar{\eta}_\pi^* + \eta_c)},
$$
where $\bar{\eta}_\pi^*$ is the maximum rapidity of $\pi^0$ 
in the C.M.F.\ given by
$$
\vspace{2mm}
\bar{\eta}_\pi^* = \eta(\bar{\gamma}_\pi^*);\ \ 
\bar{\gamma}_\pi^* = \beta_c \cosh \eta_0^*,
\eqno{\rm (A3)}
$$
see Eqs. (2) and (12a) for $\eta(\bar{\gamma}_\pi^*)$ and $\eta_0^*$
 respectively, and $\bar{\eta}_\pi^* \approx \eta_0^*$ 
for $E_0 \gg M_{p}$ ($\beta_c \approx 1$).

We have to determine 
 $p_0$ in $\tau_\theta$ given by Eq. (A1a) which corresponds to the 
average transverse momentum of $\gamma$-rays, $\bar{p}_t$, see Paper I for the
 determination of $p_0$.

We have already compared our empirical cross section
in detail in Paper I with machine data in the wide energy range,
$E_0$ = 1\,GeV $\sim$ 20\,PeV, and find the present parameterization
 reproduces nicely the data even for the energy spectrum by LHC.
 However, as presented in Section 2.3, we
 use the renormalized constant given by Eq. (A2), 
which deforms slightly the cross section used in Paper\,I in the low energy
 region $E_0$ $\lsim$ 1\,GeV, while negligible in the higher energy region.

So in Fig.\ 4a, we give again the $\gamma$-ray energy spectrum
 with the revised normalization ${\it \Theta_c}$
 in the case of $E_0$ = 0.97\,GeV, where we present both empirical ones,
 the previous one (dotted curve) and the present one (solid one).
Thus we find the previous one does not reproduce the data for 
$E_\gamma$ $\gsim$ 0.7\,GeV as naturally expected, 
while the present one reproduces well the drop due to 
the constraint in the phase space.

After Paper\,I, a new data of LHCf with $\sqrt{s} = 900$\,GeV
 is reported (Adriani et al.\ [57]), so that we additionally show
 the fitting result in Fig.\ 4b. We 
 reconfirm that the present parameterization reproduces again excellently 
the production cross section in the extremely high energy region,
 in contrast to the fitting in the very low energy region in 
 Fig.\ 4a. 

In Fig.\ 4b two production cross sections are demonstrated with
 two emission angles, one with $\bar{\theta}$
= $39\,\mu$rad (solid square) and the other with $\bar{\theta}$
= $234\,\mu$rad (open square).
One must note that our parametrization reproduces
 surprisingly well the data with the set $[\bar{N}_\gamma, \bar{p}_t]$ = 
[30.2-35.2, 188MeV/c], which are plotted onto Fig. 2 in the
 text (open and filled blue circles). 

Anyway our parameterization reproduces the experimental data
 from GeV to PeV with a simple form given by Eq. (8) 
 with Eq. (9).

\newpage

\begin{table*}
\caption{
Summary of the normalized distribution function of the ${\rm e}^\pm$-spectrum, 
$\phi(\eta_\pi, q_e)$ 
and its derivative $\phi^\dagger(\eta_\pi, q_e)$, 
resulting from charged pion decay, 
$\pi^\pm \rightarrow \mu^\pm \rightarrow \rm{e}^\pm$, 
where the muon is
 created fully polarized (left-handed for $\pi^+$ decay and
 right-handed for $\pi^-$ decay).
In this table, we use a parameter, $q_e = 2E_e/m_\mu$, and
 two rapidities, $\tilde{\eta}_\mu$ (=\,0.278) and 
${\eta}_\pi$,  corresponding to that of the muon in the pion rest frame
 and that of the pion in the L.F., respectively,  
 where also summarized together are 
 $[g_1(q),\ g_1^\dagger(q)]$ and $[g_2(q),\ g_2^\dagger(q)]$
 with $q \equiv q_e{\rm e}^{\pm \eta_\pi}$, 
${\rm e}^{\pm \tilde{\eta}_\mu}$.}
\label{table1}  
\begin{center}
{\scriptsize
\begin{tabular}{llll}\\ 
 \hline \\
  $\phi(\eta_\pi, q_e)$ & 
  $\phi^\dagger(\eta_\pi, q_e)$ & 
  \hspace{-3mm} $q_e\,\subseteq\,[q_e^-, q_e^+]$ &
  \hspace{-3mm} $[\eta_\pi, \tilde{\eta}_\mu]$ \\ \\ 
\hline
\vspace{-2mm}
\\
 $g_1(q_e \mbox{e}^{\eta_\pi}) - g_1(q_e \mbox{e}^{-\eta_\pi})$
&
 $g^\dagger_1(q_e \mbox{e}^{\eta_\pi}) + g^\dagger_1(q_e \mbox{e}^{-\eta_\pi})$
& 
 \hspace{-3mm} $[0,  \mbox{e}^{-(\eta_\pi+\tilde{\eta}_\mu)}]$
&  
 --------
\\ \\
\hspace{-2mm} $g_2(q_e \mbox{e}^{\eta_\pi}) - g_2(\mbox{e}^{-\tilde{\eta}_\mu})
+ g_1(\mbox{e}^{-\tilde{\eta}_\mu}) - g_1(q_e \mbox{e}^{-\eta_\pi})$
&
 $g^\dagger_2(q_e \mbox{e}^{\eta_\pi}) + g^\dagger_1(q_e \mbox{e}^{-\eta_\pi})$
& 
\hspace{-3mm} $[\mbox{e}^{-(\eta_\pi+\tilde{\eta}_\mu)},
  \mbox{e}^{-|\eta_\pi-\tilde{\eta}_\mu|}]$
&  
 --------
\\ \\
\hspace{-2mm} $g_2(q_e \mbox{e}^{\eta_\pi}) - g_2(q_e \mbox{e}^{-\eta_\pi})$
&
$g^\dagger_2(q_e \mbox{e}^{\eta_\pi}) + g^\dagger_2(q_e \mbox{e}^{-\eta_\pi})$
& 
\hspace{-3mm} 
$[\mbox{e}^{-|\eta_\pi-\tilde{\eta}_\mu|}, 
  \mbox{e}^{+|\eta_\pi-\tilde{\eta}_\mu|}]$
& 
\hspace{-3mm} 
$\eta_\pi < \tilde{\eta}_\mu$
\\ \\
\hspace{-2mm} $g_2(\mbox{e}^{\tilde{\eta}_\mu}) -
 g_2(\mbox{e}^{-\tilde{\eta}_\mu})
+ g_1(\mbox{e}^{-\tilde{\eta}_\mu}) - g_1(q_e \mbox{e}^{-\eta_\pi})$
&
$g^\dagger_1(q_e \mbox{e}^{-\eta_\pi})$
& 
\hspace{-3mm} 
$[\mbox{e}^{-|\eta_\pi-\tilde{\eta}_\mu|},
  \mbox{e}^{+|\eta_\pi-\tilde{\eta}_\mu|}]$
&  
\hspace{-3mm}
$\eta_\pi > \tilde{\eta}_\mu$
\\ \\
$g_2(\mbox{e}^{\tilde{\eta}_\mu}) - g_2(q_e \mbox{e}^{-\eta_\pi})$
&
$ g^\dagger_2(q_e \mbox{e}^{-\eta_\pi})$
& 
\hspace{-3mm}
$[\mbox{e}^{+|\eta_\pi-\tilde{\eta}_\mu|},
  \mbox{e}^{+(\eta_\pi+\tilde{\eta}_\mu)}]$
&  
--------
\\ \\
0 & 0 
& 
\hspace{-3mm} 
$[\mbox{e}^{+(\eta_\pi+\tilde{\eta}_\mu)},\ 
  \infty]$
&  
--------
\\ \\ 
- - - - - - - - - - - - -  
&
- - - - - - - - - - - - - 
&
- - - - - - - - - - - - - 
&  
- - - - 
\\ \\ 
$\displaystyle 
g_1(q) = g_{1, 0} G(q)$ 
& 
$\displaystyle 
g_1^\dagger(q) = g_{1, 0} G^\dagger (q)$ 
& 
$\displaystyle 
\hspace{-2mm}
{g}_{1, 0}
=  \frac{m_\pi}{m_\mu} \tilde{\gamma}_\mu
(3-\xi\tilde{\beta}_\mu)$
& 
\\ \\ 
\hspace{-2mm}
$\displaystyle
g_2(q) = g_{2, 0} \Bigl[\ln \frac{q}{2\tilde{\gamma}_\mu}
 -\frac{G(q)}{g_0}\Bigr]$ 
& 
\hspace{-2mm}
$\displaystyle 
g_2^\dagger (q) = g_{2, 0} \Bigl[ 1 
 -\frac{G^\dagger (q)}{g_0}\Bigr]$ 
& 
$\displaystyle 
\hspace{-2mm}
{g}_{2, 0}
=  \frac{m_\pi}{m_\mu}
\frac{\tilde{\gamma}_\mu(\xi + 5\tilde{\beta}_\mu)}
{6({\tilde{\gamma}_\mu}^2 -1)}$
& 
\\ \\ 
$\displaystyle
G(q) = q^2 - \frac{4}{9}q^3 \mbox{e}^{-\xi \tilde{\eta}_\mu}$
&
$\displaystyle
G^\dagger (q) = 2q^2 - \frac{4}{3}q^3 \mbox{e}^{-\xi \tilde{\eta}_\mu}$
& 
\hspace{-3mm}
$\displaystyle 
g_0 = 
\frac{2}{3} \frac{1+(2\xi-3)\tilde{\beta}_\mu}
{(1-\tilde{\beta}_\mu)\mbox{e}^{-2\tilde{\eta}_\mu}}$
\\ \\
\hline 
\\
\end{tabular}
}
\end{center}
\end{table*}

\begin{table*}
\caption{
Numerical values of coefficients appearing in the multiplicity 
 given by Equation (28) for various kinds of secondaries $\lq \lq s$".}
\label{table2}  
\begin{center}

\begin{tabular}{ccccc}\\ 
 \hline 
$\lq \lq s$"
&
$\bar{N}_0$
&
$E_{\rm th}$(GeV)
&
$\epsilon_c$(GeV)
&
$k$
\\ 
\hline
$\gamma$
& 8.50
& 0.36
& 0.021
& 1.0
\\ 
$\pi^+$
& 4.20
& 0.35
& 0.021
& 1.0
\\ 
$\pi^-$
& 3.50
& 0.76
& 0.279
& 0.5
\\ 
K$^+$
& 0.36
& 2.50
& 0.021
& 1.0
\\ 
K$^-$
& 0.26
& 15.0
& 0.279
& 1.0
\\ 
\hline 
\\
\end{tabular}
\end{center}
\end{table*}

\clearpage


\begin{figure}
\begin{center}

    \includegraphics[width=8cm]{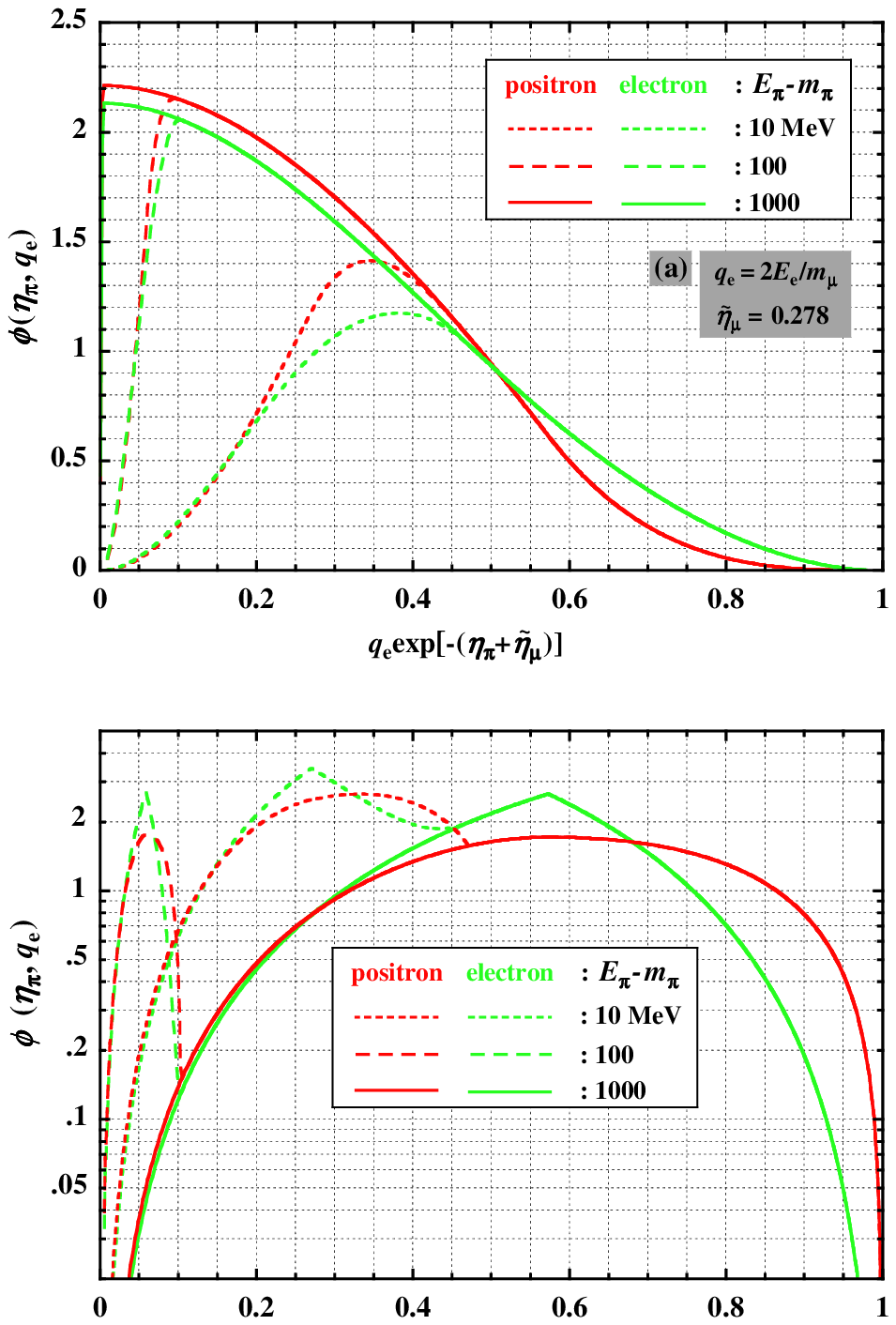}
\end{center}
  \caption{
Numerical values for (a) $\phi(\eta_\pi, q_e)$ and (b)
$\phi^\dagger(\eta_\pi, q_e)$. 
For $\epsilon_\pi (\equiv E_\pi - m_\pi)$ larger than 1\,GeV,
 both $\phi$ and $\phi^\dagger$ scale in the forms of
 $\phi(E_e/E_\pi)$ and $\phi^\dagger(E_e/E_\pi)$, respectively, 
 as $q_e\exp(-\eta_\pi) \approx E_e/E_\pi$.
}
\end{figure}

\begin{figure}[!t]
\begin{center}
   \includegraphics[width=8cm]{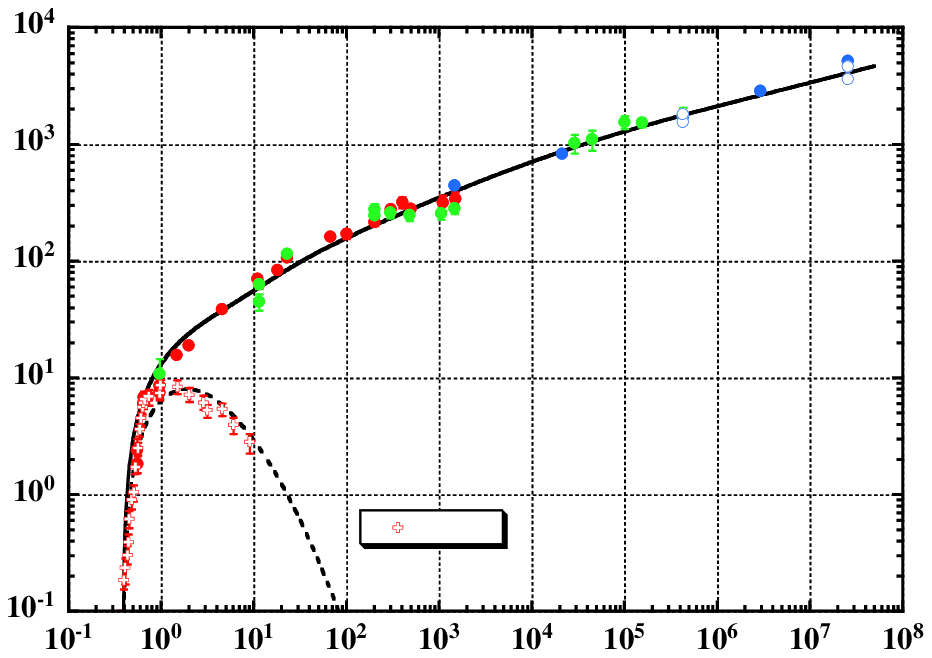}
\end{center}
  \caption{
 Total production cross sections of (a) $\gamma$-rays,
$\sigma_{\rm inel}\times (\bar{N}_\gamma + \bar{N}_\gamma^\Delta)$
 and (b) charged mesons, 
$\sigma_{\rm inel}\times (\bar{N}_{\rm ch} + \bar{N}_{\rm ch}^\Delta)$
 as a function of the L.F.\ proton kinetic energy, $E_0$.
 Broken curves correspond to the production cross sections 
  from isobaric components, and solid ones to the superposed cross
  sections with [pionization + isobaric] components. 
 See Dermer [56] for data with red symbols in both (a) and (b), and Sato et al.\  [1] for data with green in (a) and blue symbols in both (a) and (b), and 
Antinucci et al.\ [55] for  CERN-Bologna data on K$^\pm$ in (b).
}
\end{figure}

\begin{figure}
\begin{center}
    \includegraphics[width=17cm]{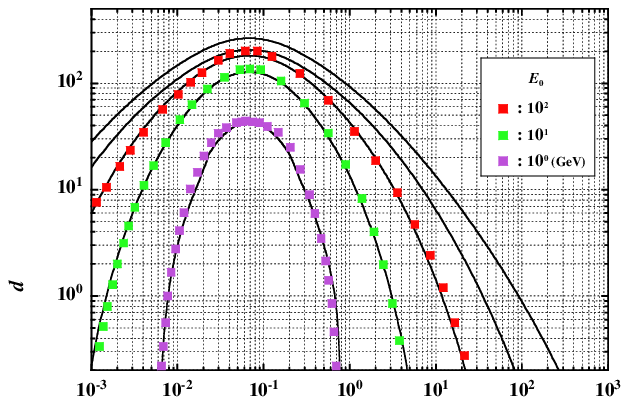}
\end{center}
  \caption{
Comparisons of the present production cross sections (heavy solid curves) for 
(a1-c1) $\sigma_{pp\rightarrow\gamma}$,
(a2-c2) $\sigma_{pp\rightarrow e^+}$, and 
(a3-c3) $\sigma_{pp\rightarrow e^-}$, with others (square symbols) by
 (a1-a3) Dermer [19], (b1-b3) Kamae et al.\ [21],
 and  (c1-c3) PYTHIA (Sjostrand et al.\ [36]), 
for $E_0$ = $10^0, 10^1, 10^2, 10^3, 10^4$\,GeV.
}
\end{figure}

\begin{figure}
\begin{center}
    \includegraphics[width=8cm]{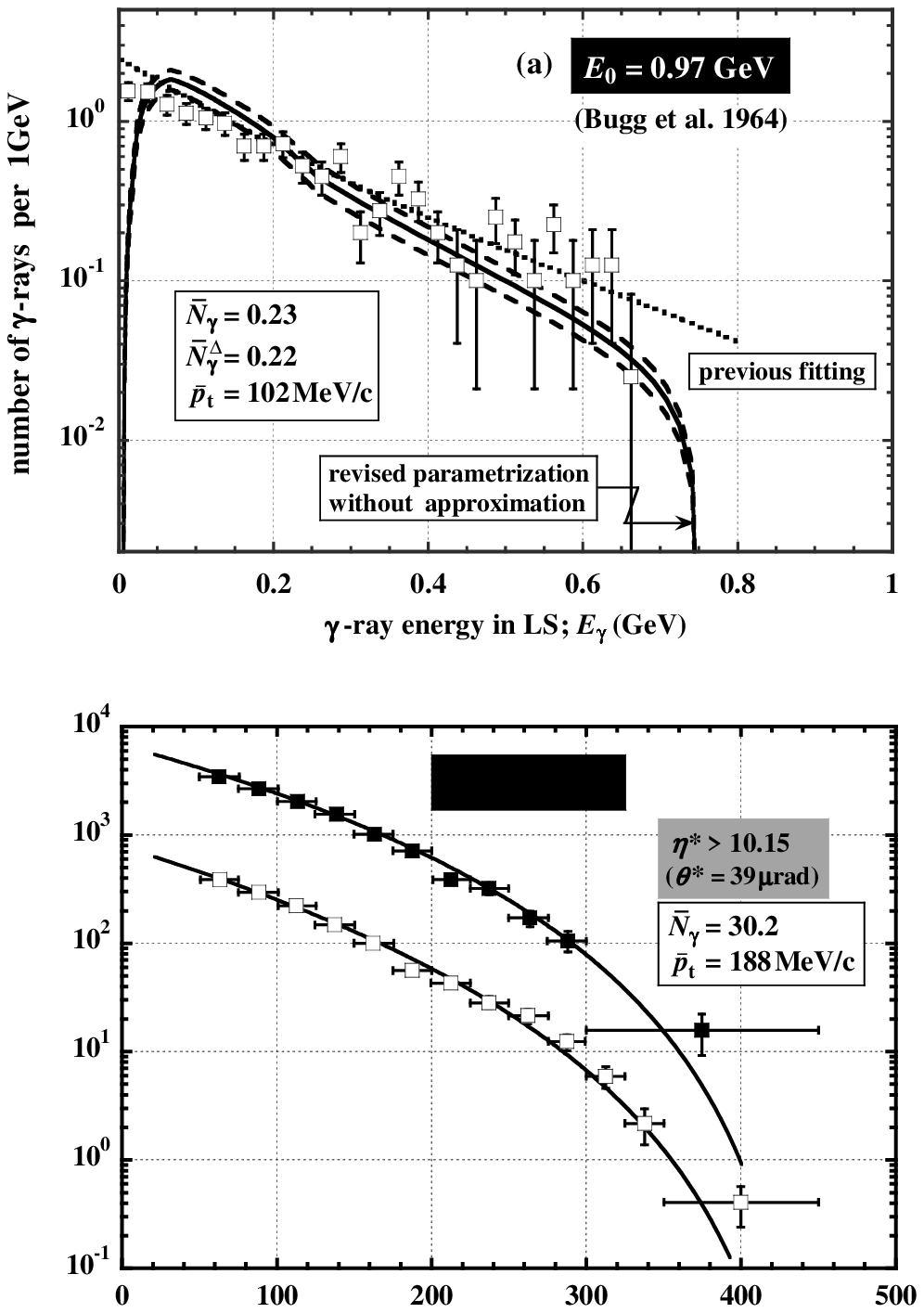}
\end{center}
  \caption{
(a) Production cross section of $\gamma$ rays at $E_0$\,=\,0.97\,GeV,
where we present empirical ones both from the old one (dotted curve)
 and the revised one (heavy solid one). (b) Production cross section of
 $\gamma$ rays at  $\sqrt{s}$\,=\,900 GeV, corresponding to $E_0 \approx$ 400
 TeV, where solid curvess are obtained by the present empirical ones.
}
\end{figure}

\end{document}